\documentclass[preprint,secnumarabic,amssymb, aps, prl]{revtex4-1}

\usepackage{natbib}
\usepackage{amssymb, latexsym}
\usepackage{amsmath} 
\usepackage{amsthm}
\usepackage{bm}
\usepackage{booktabs}
\usepackage[mathscr]{eucal}
\usepackage{enumerate}
\usepackage{url}
\usepackage{graphicx}
\usepackage{booktabs}
\usepackage{lineno}

\begin{document}

\title{Transformation seismology: composite soil lenses for steering surface elastic Rayleigh waves}

\author{Andrea Colombi}%
\email[Corresponding author ]{e-mail: andree.colombi@gmail.com}
\affiliation{Dept. of Mathematics, Imperial College London, South Kensington Campus, London}
\author{Sebastien Guenneau}
\affiliation{Institut Fresnel-CNRS (UMR 7249), Aix-Marseille Universit\'e, 13397 Marseille cedex 20, France}
\author{Philippe Roux}%
\affiliation{ISTerre, CNRS, Univ. Grenoble Alpes, France, BP 53 38041 Grenoble CEDEX 9}
\author{Richard V. Craster}
\affiliation{Dept. of Mathematics, Imperial College London, South Kensington Campus, London}
\date{\today}%

\begin{abstract}
Metamaterials are artificially structured media that can focus (lensing) or reroute (cloaking) waves, and typically this is developed for electromagnetic waves at millimetric down to nanometric scales or for acoustics or thin elastic plates at centimeter scales. Extending the concepts of \cite{kadic2013}  we show that the underlying ideas are generic across wave systems and scales by generalizing these concepts to seismic waves at frequencies, and lengthscales of the order of hundreds of meters, relevant to civil engineering.  
By applying ideas from transformation optics we can manipulate and steer Rayleigh surface wave solutions of the vector Navier equations of elastodynamics; this is unexpected as this vector system is, unlike Maxwell's electromagnetic equations, not form invariant under transformations. As a paradigm of the conformal geophysics that we are creating, we design a square arrangement of Luneburg lenses to reroute and then refocus Rayleigh waves around a building with the dual aim of protection and minimizing the effect on the wavefront (cloaking) after exiting the lenses. To show that this is practically realisable we deliberately choose to use material parameters readily available and this metalens consists of a composite soil structured with buried pillars made of softer material. The regular lattice of inclusions is homogenized to give an effective material with a radially varying velocity profile that can be directly interpreted as a lens refractive index. We develop the theory and then use full 3D time domain numerical simulations to conclusively demonstrate the validity of the transformation seismology ideas: we demonstrate, at frequencies of seismological relevance $3-10$ Hz, and for low speed sedimentary soil ($v_s: 300-500$ m/s), that the vibration of a structure is reduced by up to 6 dB at its resonance frequency. This invites experimental study and opens the way to translating much of the current metamaterial literature into that of seismic surface waves. 

\end{abstract}
\maketitle

Mathematicians and physicists have long studied the physics of waves at structured interfaces: Back in 1898, Lamb wrote a seminal paper on reflection and transmission through metallic gratings \cite{lamb98} that then inspired numerous studies on the control of surface electromagnetic waves. The concept that periodic surface variations could create and guide surface waves has emerged in a variety of guises: Rayleigh-Bloch waves for diffraction gratings \cite{wilcox78a,porter99a}, Yagi-Uda antenna theory \cite{hurd54a,sengupta59a} and even edge waves localised to coastlines \cite{evans93a} for analogous water wave systems. Most notably in the last decade, the discovery of spoof plasmon polaritons (SPPs) \cite{pendry2004} has motivated research not only in plasmonics \cite{maradudin2005,hibbins2005,maier2007} but also in the neighbouring, younger, field of platonics \cite{ross2009}, devoted to the control of flexural (Lamb) waves in structured thin elastic plates. The extension of ideas such as cloaking \cite{Pendry23062006,Leonhardt23062006,sheng2010,wegener2010} to plasmonics \cite{alu2005,paloma2010,liu2010,renger2010} is highly motivational as it suggests that one can take the concepts of wave physics from bulk media, as reviewed in \cite{fleury14a}, and apply them to surfaces. In platonics, implementations of ideas from transformation optics \cite{PhysRevLett.103.024301,wegener2012,andrea5} show how, all be it for the limiting cases of thin elastic plates, similar concepts can take place in another arena of physics, in this case mechanics and elasticity. 
Unfortunately in much of elastic wave theory, and for many applications, one actually has the opposite physical situation to that of a thin plate, that is, one has an elastic material with infinite depth; on the surface of such a half-space elastic surface waves, Rayleigh waves, exist \cite{rayleigh85a,viktorov67a,achenbach84a} that exponentially decay with depth and have much in common conceptually with surface plasmons in electromagnetism. It is therefore attractive to investigate whether concepts proven in plasmonics can be translated across despite the underlying governing equations having many fundamental differences. Some experimental work has taken place in broadly related scenarios such as the attenuation of Rayleigh waves at high frequencies in a marble quarry with cylindrical holes \cite{sanchez1999} and in piezo-electric substrates with pillars \cite{younes2011}, but with differing aims and not at frequencies relevant for seismic waves. Some work on structured elastic media, going beyond elastic plates, to try and create seismic metamaterials \cite{kim2012,brule,colombi_tree} is underway with some success either with subwavelength resonators \cite{colombi_tree} or with periodic  structuring within the soil \cite{brule} and trying to still utilise flexural wave modelling. Our aim here is complementary in that we want to implement ideas from transformation optics into this elastic surface wave area and investigate what can be achieved. 
 
The desire, and need, to control the flow of waves is common across wave physics in electromagnetic, acoustic and elastic wave systems \cite{engheta2006,sar2008,craster2012}. Acoustics is, mathematically, a simplified version of full elasticity with only compressional waves present: full elasticity has both shear and compression, with different wave speeds, and full coupling between - leading to a formidable vector system. 
The quest for a perfect flat lens \cite{pendry2,pendry02a} with unlimited resolution and an invisibility cloak \cite{Pendry23062006,Leonhardt23062006} that could conceal an object from incident light via transformation optics \cite{chen2010} are exemplars for the level of control that can, at least, in theory be achieved for light \cite{pendry15a} 
 or sound \cite{cummer2007,sanchez2008,fang2011}. An invisibility cloak for mechanical waves is also envisioned \cite{milton2006,brun2009,norris2011,norris2012} for bulk waves, and  experiments using latest developments in nanotechnology support these theoretical concepts \cite{kadic2014}. However, some form of negative refractive index appears to be necessary to achieve an almost perfect cloak. Unfortunately, a negative index  material emerges from a very complex microstructure that is not feasible for the physical length-scales, for seismic waves at frequency lower than 10 Hz this means  wavelengths of the order of a hundred meters, in geophysical applications. Notably one can design locally resonant metamaterials that  feature deeply subwavelength bandgaps \citep{colombi_tree} and so could be used in seismic and acoustic contexts, but these effects are observed over a limited frequency band; this is being improved to allow for simultaneous protection and cloaking \citep{andrea5} and has potential. However, here we  look at protection and cloaking,  applied to low frequency seismic waves, using the concept of lensing.
  
Elastic waves, in the same way as light, are subject to Snell's law of refraction, and an appropriate choice of material properties leads to spectacular effects like those created by gradient index (GRIN) lenses. Compared to classic lens where ray-paths are bent through discontinuous interfaces causing losses and aberrations, GRIN lenses are obtained with a smooth refractive index transition. Rayleigh and Maxwell themselves studied GRIN lenses, notably the Maxwell's fisheye whose spatially varying refractive index was later shown to be associated with a stereographic projection of a sphere on a plane \cite{leonhardt09a}. As noted in \cite{leonhardt09a} GRIN lenses have been mainly studied and implemented for optical applications, or wave systems governed by the Helmholtz equation. The ideas behind transformation optics are not limited to metamaterial-like applications and have contributed to recent advances in plasmonic light confinement by touching metallic nanostructures \cite{pendry2010}.
 
A classical example of a GRIN lens is the circular Luneburg lens \cite{luneburg_optic,leonhardt09a,chen2011}, once a plane wave enters the lens, its radially varying refraction index steers the ray-path towards a focal point located on the opposite side of the lens \cite{2040-8986-14-7-075705}. The eponymous fisheye lens of Maxwell \cite{maxwell1953} is another well documented and fascinating GRIN lens \cite{Smolyaninova2010} and it has been proposed as a non-Euclidean cloak \cite{leonhardt2006,tyc2009}. To date, applications have been mainly limited to scalar wave systems governed by Helmholtz's type operators for transversely polarized electromagnetic waves \cite{lens_nano,lens_nature}, for pressure waves \cite{chang2012} or platonics where composite and thickness modulated plates have recently been proposed \cite{torrent_plate2,climente,sebbah_plate,matthieu}. In full elasticity, where the propagation is described by the vector Navier equation that is not simply a Helmholtz equation or scalar, a proof of concept is still missing. 
Experimentally, for seismic waves, the realization of any proposed lensing arrangement becomes a real challenge because wavelengths range from $20$ m to $500$ m. For real applications wave control must be achieved over a broad frequency band, to cover the various wavelengths of interest; unlike other metamaterial cloak designs such as those based around subwavelength resonators \cite{andrea5}, the lens proposed here is effective across a very broad spectrum of frequencies. 

In civil engineering the structuring or reinforcement of soils is commonplace with  geotechnical solutions aimed at improving soil seismic performances \cite{ground_improvement} (e.g. soil improvement by dynamic compaction, deep mixing and jet grouting), typically implemented prior to construction of structures,  aimed to rigidify or decouple the building response and not at rerouting the seismic input. In the conceptual lens we design, the structuring of the soil is feasible and we use material parameters typical of poorly compacted sediments. Using  the ideas of transformation optics, and detailed 3D numerical simulations, we show that a square arrangement of four Luneburg lenses can completely reroute waves around an area for seismic waves coming with perpendicular incident directions (e.g. $x$ and $y$ in Fig. 1b). Not only are the waves rerouted leaving the inner region protected, but the wave-front is reconstructed coherently after leaving the arrangement of lenses. 



\section{Luneburg lens for seismic surface waves}
In his seminal work, Luneburg \citep{luneburg_optic} derived a spherical optical lens with radially varying refractive index that focused a beam of parallel rays to a point at the opposite side of the lens; a two dimensional variant is straightforward to deduce. Of course this relies upon the governing equation being the Helmholtz equation, which the full ela

In the model configuration presented here the elastic energy is primarily carried by Rayleigh surface waves; they are a particular solution of Navier's equation for elastodynamics for a half-space bounded by a traction-free surface, e.g. the Earth's surface. Well known in seismology, for the idealised situation of isotropic and homogeneous media they are non-dispersive, elliptically polarized and in practical terms \cite{viktorov67a} they have a velocity very close to that of shear waves: $v_s^2={\mu}/{\rho}$ where $\mu$ is the shear modulus and $\rho$ the density \citep{global_seism}  so for simplicity we will simply use the shear wave speed in our analysis. Shear horizontally polarized waves (SH) are also present in our numerical model, and they also propagate with wavespeed $v_s$; notably SH waves are governed by a Helmholtz equation without any approximation.
 We do not consider Love waves here, which can also be important is seismology,  as they only exist for stratified layered media and we assume that our elastic half space is vertically homogeneous, that is, the material parameters do not vary with depth. In Cartesian coordinates we take $z$ to be the depth coordinate and $x,y$ to be in the plane of the surface, then the Rayleigh waves can be represented using a Helmholtz equation on the surface and we consider a circular lens on the $x-y$ plane as in Fig. 1c, is characterized by a radially varying refraction profile \cite{2040-8986-14-7-075705}. This lens, and the associated material variation, then extends downwards and the material is considered vertical homogeneous; we distinguish the material outside the lens to have parameters with a subscript $0$ and that inside to have subscript $1$.

The refraction index $n$ between two media, say, material 0 and material 1 can be formulated in terms of the ratio of velocity contrast $n=\dfrac{v_0}{v_1}$.
For a Luneburg lens we require the refractive index, $n(r)$, to be:
\begin{equation}
\label{eq:ref_lune}
n(r)=\sqrt{2-\frac{r^2}{R^2}};
\end{equation}
where $r=\sqrt{x^2+y^2}$ is the radial coordinate and $R$ the outer radius of the lens (Fig.~1c).
We tune the material velocity within the lens to reproduce the index given in \ref{eq:ref_lune} so 
\begin{equation}
\label{eq:vel_profi}
v_{s1}=\frac{v_{s0}}{\sqrt{\left( 2-\dfrac{r^2}{R^2}\right)}}.
\end{equation}
Taking a continual material variation is perfect for theory, but from a practical perspective it is not possible to realize a circular structure 10's of meters in depth and radius, whose soil properties change smoothly (e.g. on the scale of Fig. 1c). Instead we create a composite soil made of bimaterial cells such that their effective material properties have the variation we desire, this provides a realistic lens using  actual soil parameters that could be created using conventional geotechnical techniques \cite{deep_mixing,ground_improvement}. 

In Fig.~1c the circular surface of the lens is discretized using equally spaced cells on a periodic square lattice. Each cell contains an inclusion of softer material that, in our illustration, is represented by a pillar extending down into the soil; the exponential decay of the Rayleigh wave amplitude with depth means that for the computational model we can truncate this and a depth of 30 m is more than sufficient. The diameter of each pillar is determined using the effective velocity prescribed for each cell based upon its radial position ($r$) from the center of the lens. Assuming a square section cell of width $l$ on the $x-y$ plane the filling fraction is defined using the surface area occupied by the pillar in the cell. For cylindrical pillars with diameter $d$ (Fig.~1c) we have a geometrical filling fraction, $f$, with $f=1-\dfrac{\pi d^2}{4l^2}$. The Maxwell-Garnet formula \cite{composites}, derived for composites, relates the filling fraction with the corresponding effective property:
\begin{equation}
\label{eq:garnett}
f=\frac{(v_{se}-v_{s0})(v_{s0}+2v_{ic})}{(v_{se}+2v_{ic})(v_{se}-v_{ic})};
\end{equation}
where $v_{se}$ is the effective shear velocity in the cell and $v_{ic}$ is the shear velocity of the inclusion (the pillar). We combine the geometrical definition of $f$ with (\ref{eq:garnett}) to obtain the effective velocity as a function of inclusion size. Hence, by tuning the pillar diameter we obtain the required velocity variation desired in Eq. \ref{eq:vel_profi} and use this to define the structure and variation for each of the Luneburg lenses one of which is shown in (Fig.~1c).

We now place four Luneburg lenses as shown in Fig.~1b and use these to protect an object placed in between them. The idea is simply that a plane wave incident along either the $x$ or $y$ axes will be focussed by the lens to a single point, the point at which the cylinder touches its neighbour, which will then act as source into the next Luneburg lens and the plane wave will then reemerge unscathed; the building to be protected should, in this perfect scheme, be untouched. We are aiming to demonstrate the concept not in a perfect scenario, but using realistic parameters and a setting in which the effective medium approach provide a discrete velocity profile, yet the protection achieved is considerable. 

\begin{figure}
\centering{}\includegraphics[clip,width=18cm,trim = 0mm 0mm 0mm 0mm]{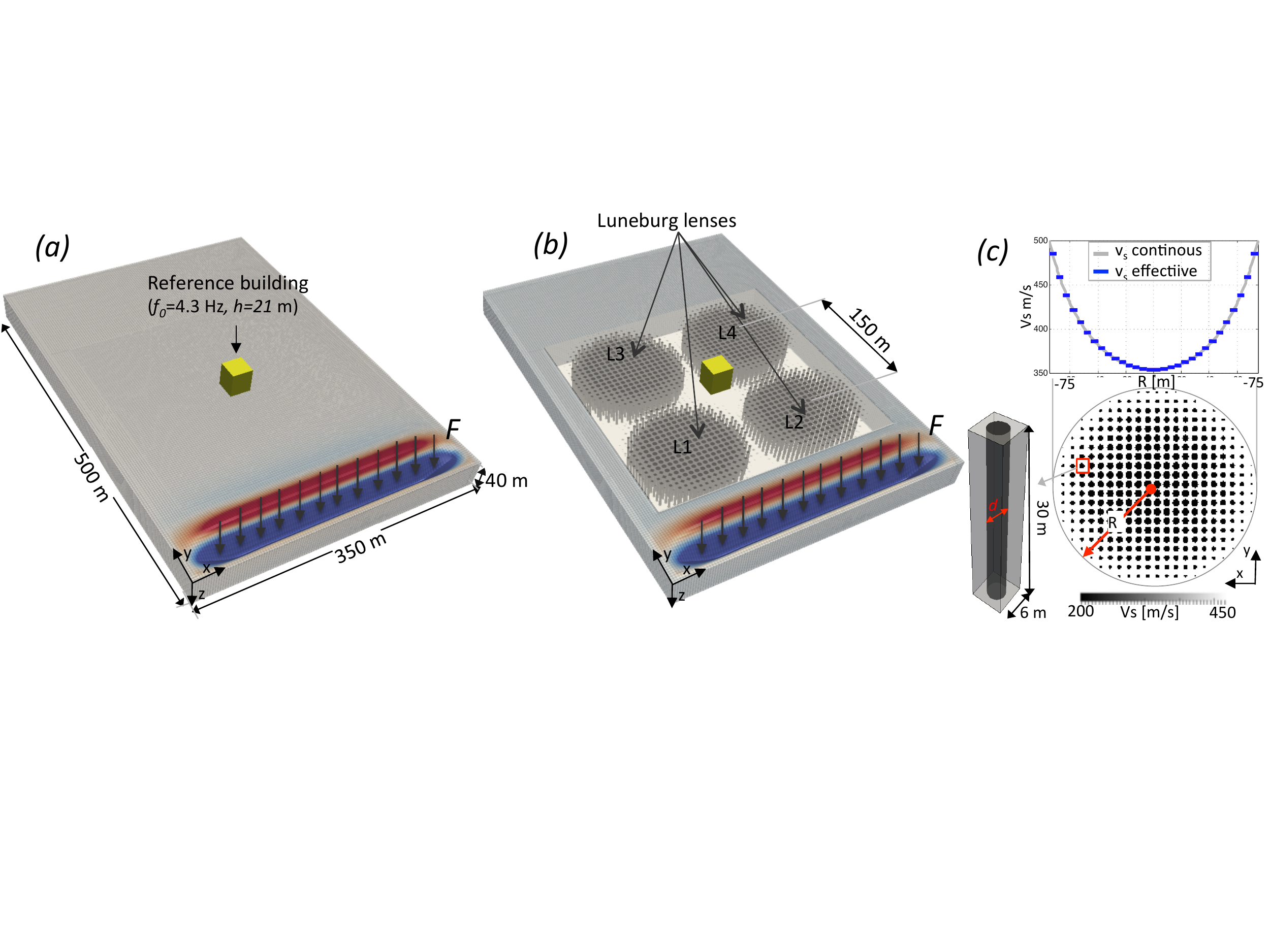}
\protect\caption{The vertical component of the displacement field is depicted at the onset time along with the meshed model of the halfspace that is used as input for the SPECFEM3D simulations. The force generating the plane surface waves is represented only in its vertical component. The building is made of stiffer material and it is also meshed with the halfspace. (b) Same as (a) but here we see the pillars forming the lens that enclose and protect the building. (c) The velocity profile as function of the radius is depicted in blue for equation (1) and in gray for the effective velocity of each cell. The inset shows a zoomed view of the cylindrical pillar and the cell. \label{fig:1}}
\end{figure}

To construct the Luneburg lenses, to reach the minimum $v_s$ prescribed in Eq. \ref{fig:1}, $v_{ic}$ needs be lower than 350 m/s. We choose a $v_{ic}$ of 200 m/s which is a value that is realistic for poorly consolidated soil (sand or water filled sediments) \citep{foti,Cornou01122003}. In the lens configuration depicted in Figs. 1b and c for each lens there are 26 elementary cells ($\sim6\times 6$ m) along the radial axis of the lens and the diameter of the pillars increases towards the center of the lens as discussed earlier. In the frequency range we investigate (3-8 Hz), the inclusion is deeply subwavelength and non-resonant. The only parameter of interest for the lens design using composite soil is the filling fraction, so there is no bound on the size of the elementary cell so long as it remains subwavelength to avoid Bragg scattering phenomena; in our simulations the minimum size of each cell is bounded for numerical reasons (explained in the next section). For an actual implementation of the lens the cell could be chosen to be even smaller than our choice here with corresponding decrease in the pillar diameters. A maximum diameter of approximately 2 m would permit the pillars to be realised with existing geotechnical machineries \citep{deep_mixing,ground_improvement}. 

\section{Numerical simulations}
Three dimensional numerical simulations of seismic elastic surface waves are implemented using SPECFEM3D: a parallelized, time domain, and spectral element solver for 3D elastodynamic problems widely used in the seismology community \citep{Komatitsch01041998,specfem_cart,earth_simu,Rietmann:2012,Magnoni01022014,GJI:GJI967}.

The reference computational domain, depicted in Fig. 1a, is a 40 m depth halfspace 350 x 500 m wide of homogeneous sedimentary material with background velocity $v_s$ set to 400 m/s. The computational elastic region, apart from the surface itself, is surrounded by perfectly matched layers (PMLs) \cite{berenger94a} that mimic infinity and prevent unwanted reflections from the computational boundaries and these are standard in elastic wave simulation \cite{skelton07a,Komatitsch_CPML}. 

A small building with a flexural fundamental mode of approximately 4 Hz is located at the center of the model. 
We place a line of broadband Ricker source time functions \citep{GJI:GJI967} centered at 5 Hz to generate an almost perfectly plane wavefront (Figs. 1a and b). The driving force {\bf F} has equal components in all 3 orthogonal directions and hence the resulting excitation is not only made of Rayleigh but also of SH waves. Body waves leave the computational domain and pass into the PML at the bottom and side boundaries of the computational domain; their interaction with the structure and the lens is negligible. This configuration could represent the upper layer of a deep sedimentary basin with some strategic structure located at the surface (power plants, data centers, hospitals) desired to be shielded from a seismic energy source. 
In Fig. 1b representing the building surrounded by the square array of lenses, the pillars are inserted as softer inclusions in the background velocity model. This approach is commonly used \citep[e.g.][]{Lee201456,Magnoni01022014} to simplify the discretization of complex models. We use very fine meshing (down to 2 m) and this, combined with $5^{th}$ order accuracy in space of the spectral element method, allows us to accurately model pillars down to a smallest diameter of 0.3 m. 
This approach was validated against a model where the pillars were meshed explicitly; the only notable difference was a factor of 5 increase in the runtime for the explicit case vis-a-vis the regular mesh.
SPECFEM3D is a parallel code and simulations are run on 64 cores for a total of approx. 30 corehours for a simulated time of 1.5 seconds with the 3D wavefield saved for post-processing; SPECFEM3D is the standard geophysics code used in academic and industry applications with a long history of development and application \citep{Komatitsch01041998,specfem_cart,earth_simu,Rietmann:2012,Magnoni01022014,GJI:GJI967}. 

\section{Results}
This simulation is shown for wave-field snapshots for different propagation distances in Fig. 2 both with, and without the lenses. The sources generating the plane wavefront in Fig.~1 are located at the surface and so most of the seismic energy propagates as Rayleigh and SH waves. The vertical component of the displacement $\mathbf{u}$ shown in Fig. 2, is dominated by the elliptically polarized motion of the Rayleigh waves. Although not visible, SH waves behave very similarly to Rayleigh waves for the model here discussed, body waves have far lower amplitude and are not relevant in our analysis. Fig. 2a shows, as one would expect, the strong scattering by the building and its highly oscillatory motion. 
 When the Luneburg lenses are inserted in the model (Fig. 2b) the simulation shows that the Rayleigh wave front splits and then progressively converges to the focal points of lenses L1 and L2. Given the square layout, the focal points lie exactly at the touching points of the L1-L3 and L2-L4 lenses. This  lens does not support any subwavelength phenomena (the evanescent part is lost) hence, the size of the focal spot is diffraction limited at $\lambda/2n$ and some energy is backscattered during the focusing process creating some low amplitude reflections. The second lenses (L3 and L4) behave in a reciprocal manner converting the two point (secondary) source-like wavefields back into a plane wavefront. During the entire process, the inner region where the building is placed has experienced a very low seismic excitation as compared to the reference unprotected case. Fig. 3 presents the motion of the roof of the reference building on a dB scale and it shows the vibration is drastically reduced. The snapshots in the bottom row of Fig.~2 showing the wavefront as it emerges from the lenses shows that despite the strong alteration of the ray-path, the reconstruction of the wavefront after the lenses is surprisingly good. Hence this device combines the some cloaking behaviour with the seismic protection. Considering the broad frequency bandwidth of the input signal this is an interesting result as most cloaks so far proposed have problems with broadband excitation. The velocity structure of the lenses is such that the propagation of the wavefront in Fig. 2b is slightly slower than the reference configuration of Fig. 2a. Thus, we observe cloaking functionality to be valid for the wave envelope but not for the phase. This is not particularly relevant in the present seismic context where the only application of cloaking is to avoid very directive scattering, it would be unfortunate and undesirable to scatter or refocus the signal to a neighbouring building, while simultaneously realising seismic protection. 
     
A quantitative analysis of the wavefield is presented in Figs.~3a and 3b which show the energy maps for the reference and protected cases. The energy is calculated at the surface ($z=0$) taking the $L_2$ norm of the three components of the displacement field $\mathbf{u}(x,y,0)$. In the homogeneous case (Fig.~3a) of the unprotected building the energy is almost uniform across the whole computational domain making the building resonate. In the protected case, the energy is focused towards the axes of symmetry of the lenses, leaving a central region relatively undisturbed; in the two stripes shown in Fig. 3b the energy (and equally the amplitude) is much higher than elsewhere.

Fig. 3c shows the frequency response function of the building with, and without, the lenses. The rooftop horizontal displacement (roof drift) \citep{stru_dyn} is a diagnostic of the amplification phenomena due to the resonance frequency of the structure. Over the whole spectrum an average amplitude reduction of 6 dB is achieved which is reduction of almost an order of magnitude in the vibration. Complete  cancellation of the wavefield is not achieved primarily because 
the evanescent field slightly couples with the building and as we focus the wavefield to a point source we introduce some back scattering that also interacts with the building (Fig. 2b).  Nonetheless the concept is demonstrated and shows that one can successfully translate ideas from optics into the field of elastic seismic waves; Figs. 2 and 3 should inspire the use of these concepts to both reroute surface waves and to reduce their impact on surface structures.
          
\section{Concluding remarks}
We have combined concepts from transformation optics and plasmonics, composite media, and elastic wave theory to create an arrangement of seismic Luneburg lenses that can reroute and reconstruct plane seismic surface waves around a region that must remain protected. The lens is made with a composite soil obtained with columns of {\it softer} soil material with varying diameter distributed on a regular lattice. The use of softer soil inclusions emphasises that the methodology we propose is not reflection of waves, or absorption, or damping for which rigid or viscoelastic columns might be more intuitive; the softer inclusions are designed to progressively alter the material itself so that waves are ``steered'' and the reconstruction of the wavefronts after exiting the arrangement illustrates the mechanism. The Luneberg lens arrangement proposed here could be tested in a small scale experiment using elastic materials, such as metals, as the concept itself is very versatile or on larger scale test areas where one could then evaluate effects such as nonlinearity. Although presented in the context of seismic engineering there are everyday ground vibration topics that could benefit from this design. The damping of anthropic vibration sources (e.g. train-lines, subway, heavy industry) is very important for high precision manufacturing process, to reduce structural damages due to fatigue \citep{takemiya2005environmental} or simply to decrease domestic or commercial building vibrations.

Our aim here has been to present the concept of steering elastic surface waves completely in context and with a design that could be built, this should motivate experiments, further design and the implementation of ideas now widely appreciated in electromagnetism and acoustics to this field. One important practical point regarding our design is that we have presented normal incidence to the four Luneberg lens and this is practical, of course, if the position of the vibration (a railway line for instance) is known. The protection progressively deteriorates as the incidence angle of the plane wave increases; at $45^{\circ}$, the focal point is in the centre of the region (Fig.~4) and the energy is steered to this point.
The concept of using soil mediation to steer surface elastic waves is clear and the Luneburg arrangement is a clear exemplar of the ideas. 
The proposed design is easily adapted to higher frequency bands and smaller regions. If one is only interested in seismic protection, and less in the wavefront reconstruction,  the lenses can be structured differently with only one or two lenses. The other well-known GRIN lenses offer different extents and types of wave control (e.g. Maxwell, Eaton \cite{eaton1952,eaton1953}) can provide an isotropic wave shielding.
Other types of GRIN lenses and layout (for instance using 4 half-Maxwell lenses as shown in optics \cite{2040-8986-14-7-075705}) can be utilised; however the Luneburg lens requires the lowest velocity contrast between the lens and exterior region and we choose to use it as practically it could be implemented. 
The main practical difficulty is that these lenses are either singular (an Eaton lens has $v_s=0$ m/s in the center) or prescribe stronger velocity contrast (Maxwell) requiring difficult (or not yet available) soil engineering solutions.

\vspace{-1cm}
\section{Acknowledgements}
We are particularly grateful to the SPECFEM3D development team led by Dimitri Komatitsch, who helped us with the implementation of the PML conditions essential in these simulations. The authors also thank the EPSRC, ERC and CNRS for their financial support. AC was supported by a Marie Curie Fellowship.

%

\newpage

\begin{figure}
\centering{}\includegraphics[clip,width=18cm,trim = 0mm 0mm 0mm 0mm]{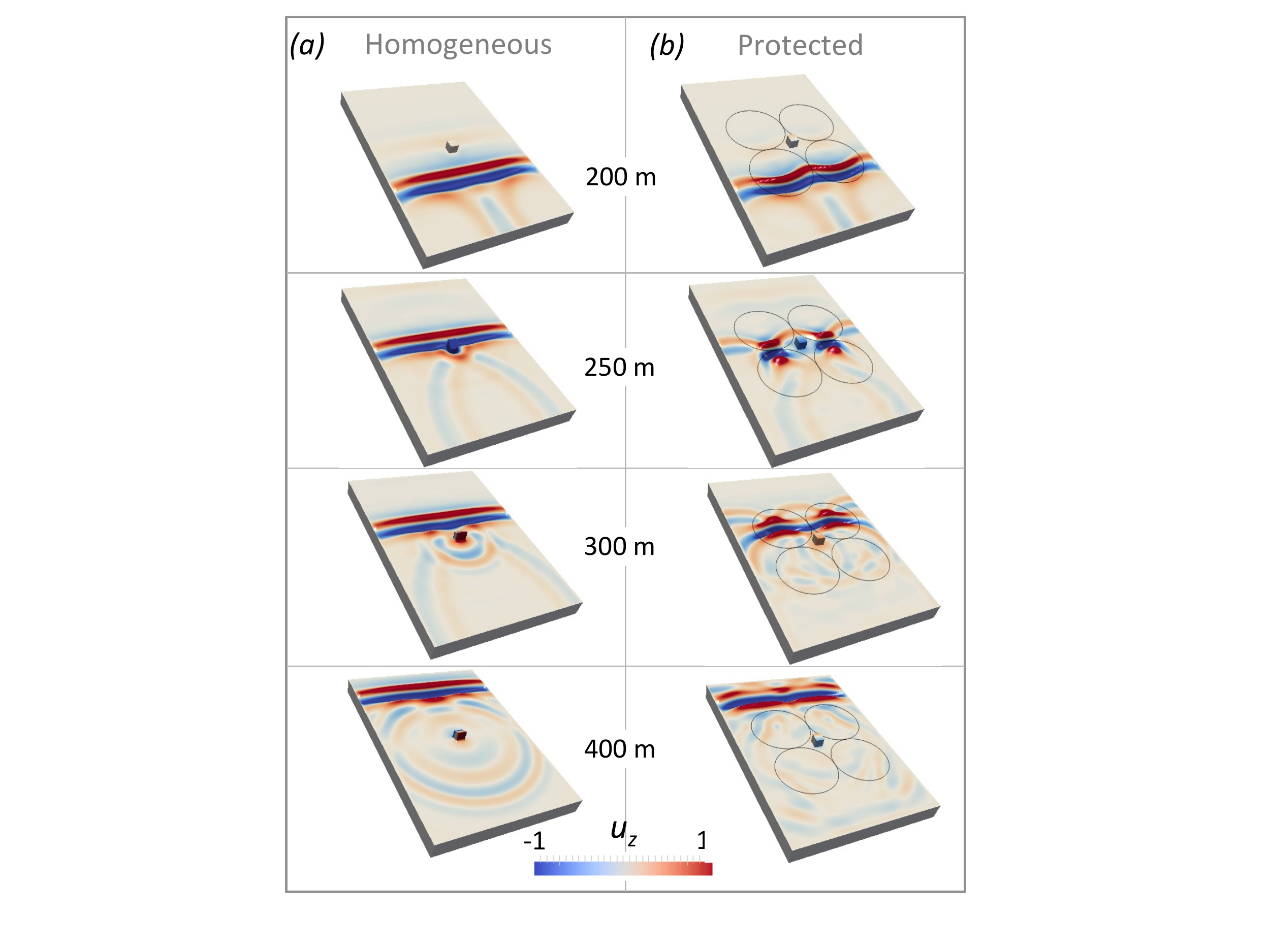}
\protect\caption{(a) Snapshot of the vertical component of the wavefield for the homogeneous case for different propagation distances from the source. The colorscale is saturated. (b) Same as (a) but with the lenses present. Full video available at (add link).\label{fig:2}}
\end{figure}

\begin{figure}
\centering{}\includegraphics[clip,width=18cm,trim = 0mm 0mm 0mm 0mm]{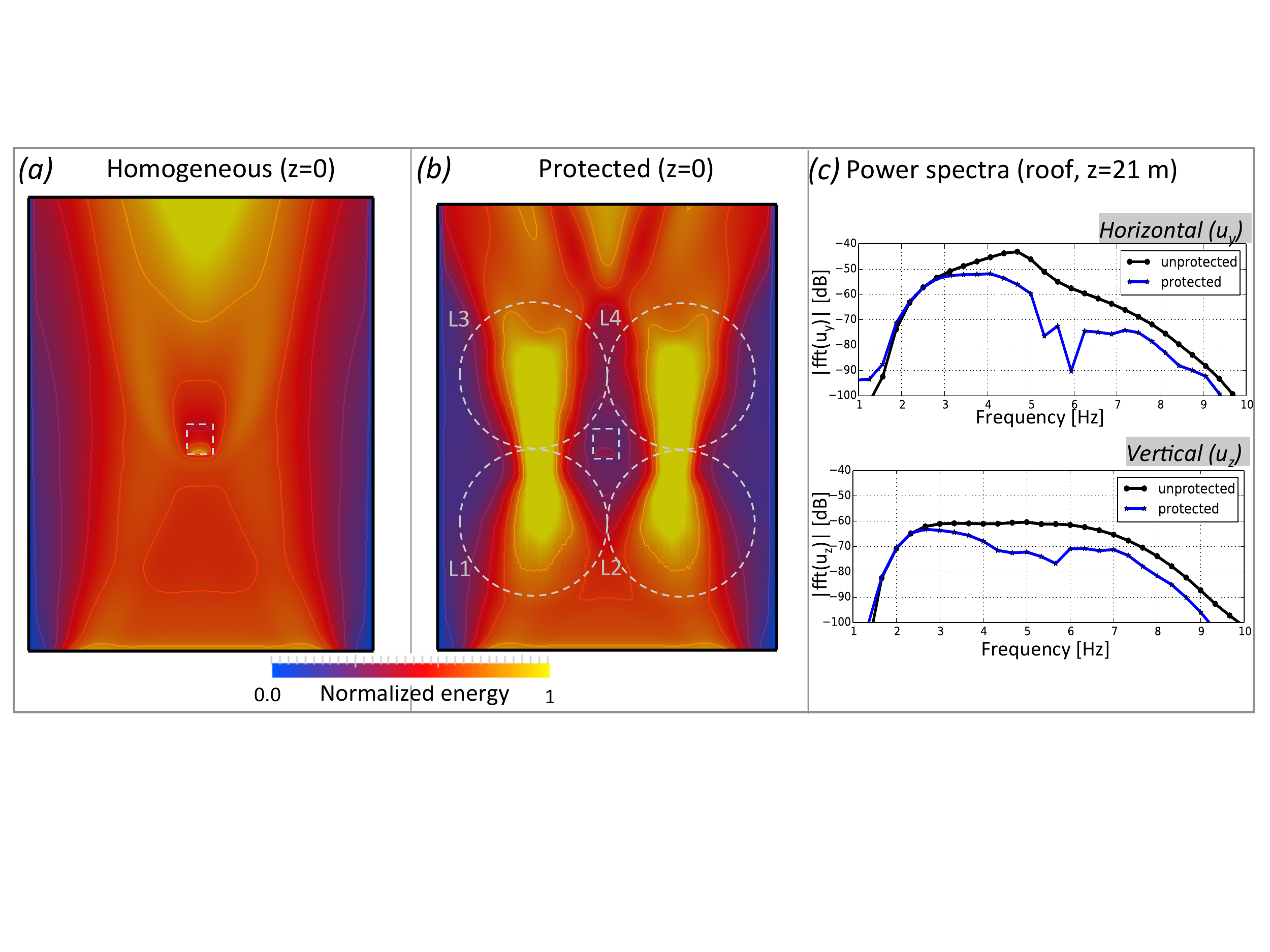}
\protect\caption{(a) Maps of the elastic energy distribution at the surface (z=0) for the homogenous case. Same as (b) but for the lenses case. (c) Spectral density of the rooftop motion of the building in dB. The blue trace is calculated for the homogeneous case while the black is obtained when the lenses enclose the building. (d) same as a (c) but for the vertical component.\label{fig:3}}
\end{figure}

\begin{figure}
\centering{}\includegraphics[clip,width=18cm,trim = 0mm 0mm 0mm 0mm]{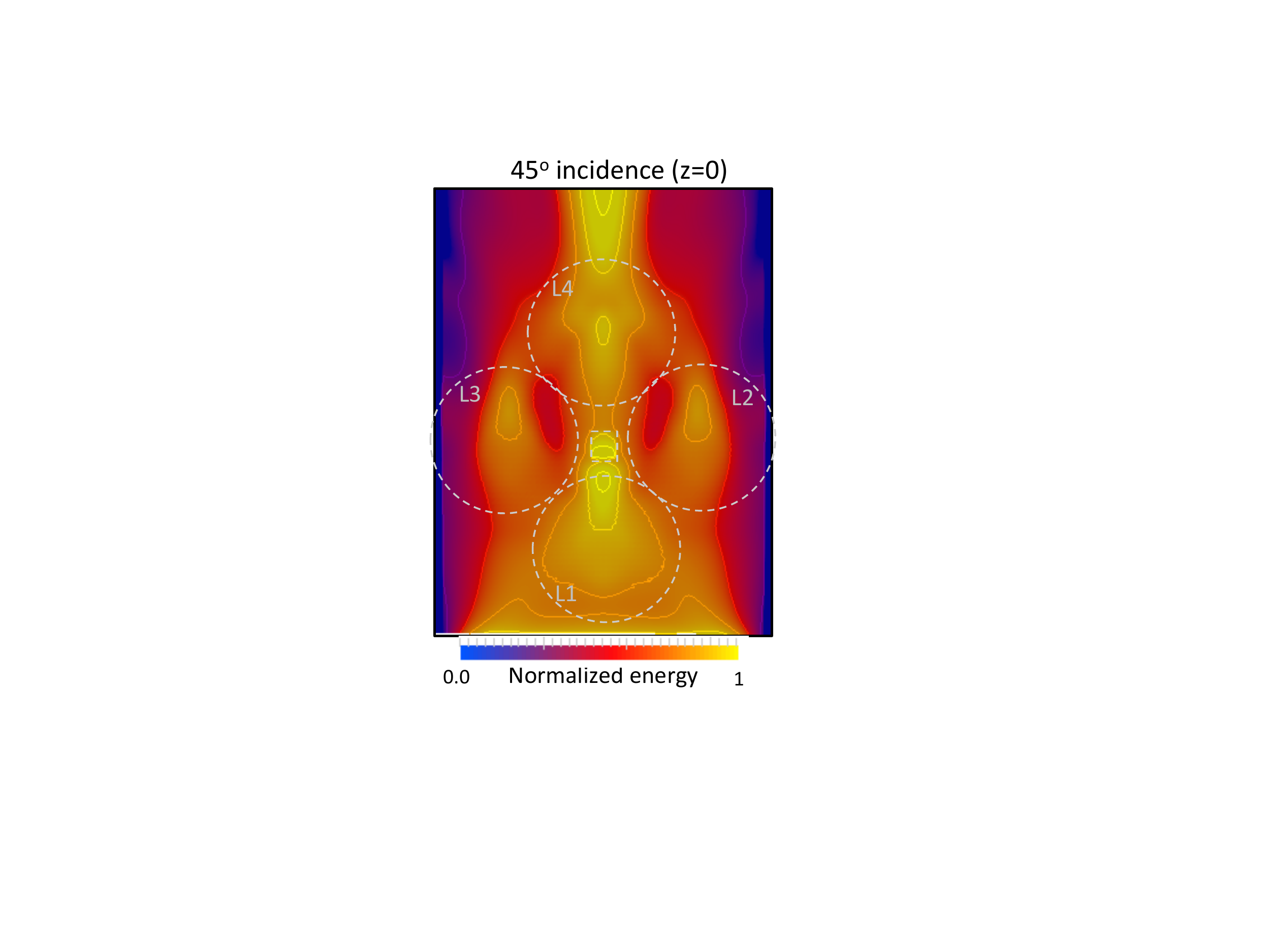}
\protect\caption{ Same as Fig. 3b but for a plane wave approaching the lenses with a 45$^{\circ}$ incidence angle. \label{fig:4}}
\end{figure}

\end{document}